%
%
\magnification=\magstep1

\font\ninerm=cmr9
\advance\mathsurround1pt
\font\ninerm=cmr9
\headline={\it J. J. Lodder, Preprint. \hfill
hep-ph/yymmddd
\hfill
Submitted to \rm  Phys. Lett. B}

\centerline{\bf\uppercase{On gluon masses, colour symmetry,}}
\medskip
\centerline{\bf\uppercase{and a possible massive ninth gluon}}
\bigskip
\centerline{ J. J. LODDER}
\smallskip
\centerline{\it Oudegracht 331\thinspace b,
3511\thinspace\thinspace PC Utrecht,
The Netherlands.}
\bigskip\bigskip\bigskip
{\ninerm\noindent
{\bf Abstract}
Gluons acquire a determinate mass from quark loop radiative corrrections
and self-interaction,
when these are recomputed with the symmetrical theory of generalised functions.
The colour symmetry may be~$U(3)$ instead of~$SU(3)$.
The symmetry breaks spontaneously and the ninth colour~$SU(3)$ singlet gluon
acquires a very large mass.
There may be some experimental support for its existence.
}
%
\begingroup
\def\sss{\scriptscriptstyle}
\newskip\diaskip \diaskip=9mm
%
\newfam\bfsyfam                                
     
\font\tenbfsy=cmbsy10
\font\sevenbfsy   =cmsy7
\font\fivebfsy    =cmsy5
\textfont\bfsyfam          =\tenbfsy
\scriptfont\bfsyfam        =\sevenbfsy
\scriptscriptfont\bfsyfam  =\fivebfsy
\skewchar\tenbfsy='60 \skewchar\sevenbfsy='60 \skewchar\fivebfsy='60
\newfam\mitbfam                               
\def\mitb{\fam\mitbfam}
\font\tenmitb   =cmmib10                      
\font\sevenmitb =cmmi7
\font\fivemitb  =cmmi5
\textfont\mitbfam          =\tenmitb
\scriptfont\mitbfam        =\sevenmitb
\scriptscriptfont\mitbfam  =\fivemitb
\skewchar\tenmitb='177  \skewchar\sevenmitb='177  \skewchar\fivemitb='177
%
\catcode`@=11                            
\def\hexnumber@#1{\ifnum#1<10 \number#1\else
 \ifnum#1=10 A\else\ifnum#1=11 B\else\ifnum#1=12 C\else
 \ifnum#1=13 D\else\ifnum#1=14 E\else\ifnum#1=15 F\fi\fi\fi\fi\fi\fi\fi}
\def\bffam@   {\hexnumber@  \bffam}      
\def\mitbfam@ {\hexnumber@  \mitbfam}
\def\bfsyfam@ {\hexnumber@  \bfsyfam}    
\def\msymfam@ {\hexnumber@  \msymfam}
\mathchardef\taubf  = "0\mitbfam@1C
\mathchardef\cdotbf = "2\bfsyfam@01
\catcode`@=12
\def\lover#1/#2 {{\textstyle{#1\over#2}}}
\def\eq(#1){(#1)}
\def\ref#1{ [#1]}
\def\intdppi{\int\!\!{d^{\,4}p\over(2\pi)^4}\,\,\,}
\def\intdp{\int\!\!d^{\,4}p\,\,\,}
\def\msq{m^2}
\def\mt{m_{t}^{\vphantom2}}
\def\mtsq{m_{t}^{\,2}}

\def\mnul{m_{\sss0}^{\vphantom2}}

\def\mgsq{m_{g}^{\,2}}
\def\psq{p^2}
\def\phat{\hbox{$\not\mkern-3.3mu p\mkern3.3mu$}}
\def\gs{g_s}
\def\gssq{\gs{}^{\!\!2}}
\def\gnul{{g_{\sss0}^{\vphantom0}}}

\def\pimm{{\Pi}_\mu^{\,\mu}(0)}
\def\as{\alpha_s}
\def\uth{\hbox{$U(3)$}}
\def\suth{\hbox{$SU(3)$}}
\def\sutw{\hbox{$SU(2)$}}

\def\gev{\hbox{\rm  GeV}}

\bigskip
\noindent{\bf 1. Introduction}
\bigskip
A new theory of generalised functions has been constructed\ref{1,2}.
The available simple model allows the multiplication
of all generalised functions needed in  quantum field theory.
Standard concepts of analysis, such as limit,
derivative,
and integral,
have to be extended to make multiplication of generalised functions possible.

Integration of symmetrical generalised functions between arbitrary limits
is always possible and yields well-defined finite results.
Infinity of integrals is replaced by the less restricted concept of determinacy,
which is related to the scale transformation properties of the integrand.
In contrast to all regularization schemes the results
{\it are not arbitrary by finite renormalizations.}

Conversely, all results in quantum field theory that depend on the use
of {\it any particular method of regularization}\/ are invalid
by the standard of the symmetrical theory of generalised functions.
In particular,
every result that is dependent on the use of dimensional regularization
to obtain it,
disagrees with the corresponding generalised function result\ref3.
This is not merely a mathematical subtlety,
it has physical consequences for observable quantities\ref{4,5}.

Renormalizability is no longer a relevant criterion for quantum field theories,
and gauge invariance may be broken dynamically.

The starting point is the strong sector of the usual standard model Lagrangian.
The~$\sutw$ symmetry is broken by hand,
by giving the top quark a non-zero mass,
leaving the bottom quark (effectively) massless.

\noindent
Note:
It is inevitable that one fermion mass is introduced to set the mass scale.
Gauge boson masses can result from radiative corrections,
conversely fermion masses cannot be generated
by interaction with massive gauge bosons.

Since only the mass terms have to be computed
it is not necessary to consider ghosts.
We can compute conveniently in the unitary gauge,
which is free of unphysical  fields.
The calculations are very similar to the corresponding calculations\ref5
in the electro-weak sector,
so only an outline is given.

\bigskip\goodbreak
\noindent{\bf 2. Quark contributions}
\bigskip

\noindent
The fermionic  contribution to the gluon self-energy is
found from the fundamental two gluon-quark vertex and the
corresponding loop diagram,
which is needed for the present purpose only at gluon momentum~$k=0$.
The gluons have pure vector coupling,
so we obtain the usual\ref2 photon mass integral,
$$g\,\cdotbf\,{\bar t\atop t}\,\cdotbf\, g\quad =\Pi_{\mu\nu}(0)=
-3{\gssq\over4}\mathop{\hbox{\rm Tr}}\intdppi
{\mitb\lambda}^\alpha{\mitb\lambda}^\beta
{\gamma_\mu
(\phat+\mt)\gamma_\nu
(\phat+\mt)\over (p^2-\mtsq)^2},\eqno{(1)}$$
multiplied by an additional factor three for summing over the quark colours.
After contracting with~$g^{\mu\nu}$ and evaluating the trace,
both over the~\suth\ generators and~$\gamma$-matrices,
one obtains
$$\pimm = {3\as\delta^{\alpha\beta}\over\pi^3}
\intdp{\psq-2\mtsq\over
(\psq-\mtsq)^2},
\eqno{(2)}$$
with~$\as=\gssq/4\pi$ as usual.
This corresponds to the photon mass integral found before\ref5.
The imaginary part of the  integral equals~$i\pi^2$, so
$$\Delta\mgsq = {1\over4i}\pimm = {3\as\over4\pi}\mtsq .\eqno{(3)}$$
The result can not be interpreted as a quantitative prediction,
since it is unclear what value of the strong coupling constant
one should use in an integral over all energies.
This point is discussed below.

The gluons acquire a large mass by interacting with the top quark.
Consequently they
now also have a self-interaction contributing to their mass.

\bigskip
\noindent{\bf 3. Gluon self-interaction}
\bigskip
\noindent
The cubic and quartic terms in the free field part of the Lagrangian
give rise to gluon self-interaction.
As long as the gluons are
massless these terms cannot contribute a mass term by self-interaction.
As soon as the gluons have acquired a mass
(by any mechanism) there will be
self-mass corrections from the mass term in the gluon propagator.

There are two Feynman diagrams contributing to the self-energy,
the loop diagram obtained from the three-gluon vertex,
and the bubble diagram obtained from the four-gluon vertex.

In the loop diagram the~$6^{\rm th}\!$ degree terms
involving~$p^4p_\mu p_\nu $ cancel,
leaving upon contraction with~$g^{\mu\nu}$ the terms
$$g\cdotbf{g\atop g}\cdotbf g\quad={\pimm}_{\rm\sss loop}^{\phantom2}=
{}-3{\gssq\over\mgsq}f^{\alpha\gamma\delta}f^{\beta\gamma\delta}
\intdppi{p^4-3\psq\msq\over(\psq-\msq)^2},\eqno{(4)}$$
taking a symmetry factor~$1/2!$ into account.
The integral taken by itself is indeterminate.

The 4-gluon vertex gives the bubble diagram,
which yields upon substitution of a gluon propagator
$${g\,\,g\atop g\,\,\,\,\,\cdotbf\,\,\,\,\, g}\quad={\pimm}_{\sss\rm
bubble}^{\phantom2}= {}+3{\gssq\over\mgsq}
f^{\alpha\gamma\delta}f^{\beta\gamma\delta}
\intdppi{\psq-4\mgsq\over\psq-\mgsq},\eqno{(5)}$$
again with a symmetry factor~$1/2!$ included.

For the~\suth\ structure constants we
have\ref6~$f^{\alpha\gamma\delta}f^{\beta\gamma\delta}=3$.
The quartic terms cancel in the sum of the two diagrams,
so the total gluon self-mass correction is the sum of (4) and (5)
$${\Delta\mgsq}_{\sss\rm self}=
{1\over4i} {\pimm}_{\rm\sss total}^{\phantom2}=
{}-{3\gssq\over 32\pi^2}
\intdp{\psq-2\mgsq\over(\psq-\mgsq)^2},
\eqno{(6)}$$
which is again proportional to the same
determinate quadratic integral~\eq(2) found previously.
The total self-mass correction for the gluons is
$$
{\Delta\mgsq}_{\sss\rm self}
= -{9\alpha_s\over8\pi}\mgsq.\eqno{(7)}$$
It is not clear what value one should take for the strong coupling
constant~$\alpha_s$ in~\eq(7),
but the pre-factor can be of order one.
The gluon self-mass correction is of the same order of
magnitude as the gluon mass itself.
Whatever mass the gluons may acquire by any mechanism
is largely annihilated again by
self-interaction.
This large self-interaction may perhaps justify the usual treatment
of gluons as massless particles.
Perturbation theory is clearly not adequate in this strongly non-linear case.
Quark and gluon contributions should not be considered separately.
The problem cannot be resolved without
an adequate understanding of confinement and the origin of the masses,
which is at present lacking.

Nevertheless the result may be taken as a qualitative indication
of what to expect from a more adequate treatment.

\bigskip\noindent
{\bf 4. A ninth singlet gluon?}
\bigskip
\noindent
Traditionally the colour symmetry group has been taken to be~\suth\
instead of~\uth.
There are only eight \suth-octet gluons~$g_1\cdots g_8$ carrying colour charge.
The ninth gluon, $\gnul$,
being the colour singlet generated by the identity matrix with colour factor
$$
\gnul={1\over\sqrt3}(r\bar r+ g\bar g+b\bar b),\eqno{(8)}$$
would not be confined by the colour force,
and it could be exchanged directly between nucleons,
which also are colour singlets.

As long as the gluons are (forced to be) massless
there are good reasons\ref7 for the choice of~$\suth$ rather than~$\uth$.
The singlet gluon~$\gnul$ would carry a strong long range force between
nucleons,
in obvious disagreement with the existence of the world as we know it.
This argument no longer holds when the ninth gluon is sufficiently massive.
Yukawa shielding will destroy the long range intertaction.

The self-mass correction (7) applies for the traditional~\suth\ octet
gluons only.
When the symmetry group is~\uth\ the additional generator is the identity
matrix.
It always commutes,
and the additional structure constants~$f^{0\alpha\beta}=0$ are all zero.
The ninth gluon has neither self-interaction nor interaction with the octet
gluons.
It retains the full quark generated gluon mass,
renamed to~$\mnul$.
The~\uth\ symmetry is spontaneously broken,
with the singlet much heavier than the octet.

Once broken~$\suth$ is accepted
we can also regard the singlet gluon mass as a free parameter,
to be measured in suitable experiments.
Perturbation theory is probably inadequate to give
a quantitative prediction in this strongly interacting case.
Moreover, our understanding of the strong interaction is not adequate at present
to exclude other possible sources of singlet gluon mass.

The singlet gluon has some properties in common
with the proposed\ref{8,9} very heavy $Z'$-boson,
which would also couple to quarks
and be relevant to the same experimental data.
Having one eliminates the need for the other.
The singlet gluon involves less change to the structure of the standard model.

\bigskip\noindent
{\bf 6. Indications from experiment}
\bigskip
\noindent
The current understanding of the standard model of the strong interaction is
believed to be in agreement with experiment.
Yet we cannot be completely sure that an additional singlet gluon
with a mass of tens of~\gev\ would spoil the agreement,
since the experiments have been fitted selfconsistently
to complicated higher order calculations.

A massive ninth gluon will cause a very short range Yukawa interaction
between quarks.
It will not cause additional quark confinement.
The Coulomb-like part of the Yukawa potential
will cause additional hard scattering.
Experimentally a massive ninth gluon will
manifest itself by an  increase in the strength of the hard,
Coulomb-like part of the strong force by a factor 9/8
at  quark energies  well above the singlet gluon mass.

It might be thought that the massive singlet gluon should be easily observable
as a resonance in quark/antiquark scattering.
This is true, but only marginally so.

In the first place the singlet effects are masked by the octet gluons.
Then the singlet gluon resonance must be very broad.
The singlet gluon will decay strongly to quark/antiquark pairs,
broadening and lowering the resonance.
The decay width to quark pairs will to be given by (~$N_f$ flavours,
quark mass ignored)
$$\Gamma_{\!\gnul}=\lover1/12 N_f\alpha_s\mnul\approx\mnul/20,\eqno{(9)}$$
with~$\alpha_s=\alpha_s(\mnul)$ the effective strong coupling constant at the
singlet mass. This will make the~$\gnul$ the shortest lived particle ever,
with a lifetime of order~$10^{-25}\cdots10^{-26}$s,
and a range before decay much shorter than the proton radius.

The already broad  resonance will be further broadened by being folded
with the parton distribution functions inside the proton.
It remains to be seen whether the proton scattering data can be fitted,
with recomputed parton distribution functions,
to strong interaction including the ninth gluon.
The recent claim\ref{10} of the possible discovery of quark constituents,
based on an observed excess of hard events at very high energy,
may actually be an indication of the existence of a massive gluon.
If this is indeed the case the mass of the singlet gluon must be at least
300~\gev,
assuming half the momentum of the proton is carried by the quarks.
This is much higher than the second order perturbation theory\eq(3) result.

The experimental errors at the highest available energy are very large,
and the interpretation of the data
is based selfconsistent fitting of parameters,
so a firm conclusion cannot be drawn at present.
Moreover,
the fitting has been done on basis of formulae ignoring singlet gluons,
which may introduce a bias.
More accurate data at higher energy beyond the singlet resonance,
and reinterpretation of the data should settle this point.
At present the interpretation of the data is too controversial
to allow conclusions to be drawn.

\goodbreak

\bigskip\noindent
{\bf 7. Conclusions}
\bigskip

\divide\parindent by 2
\item1
In second order gluons acquire a determinate mass
by interaction with the top quark.
This mass will be greatly reduced by self-interaction
for the usual~$\suth$ octet gluons.
\item2
The colour
symmetry of the strong interaction may be
(spontaneously broken)~$\uth$ instead of~$\suth$.
\item3
The ninth gluon will be a very massive colour singlet
without gluon self-interaction.
Second order perturbation theory cannot be trusted
to predict gluon masses quantitatively.
\item4
Once the ban on gluon masses is lifted,
the existence of a singlet gluon and its mass becomes
a matter to be settled by experiment.
\item5
The existence of a sufficiently massive singlet gluon
does not immediately conflict with experiment,
and may even be confirmed by the highest energy data at FNL.

\endgroup

\vfill\eject

\noindent{\bf References}
\medskip
\begingroup
\parindent=-2em
\leftskip=2em
\def\ref#1{\par\leavevmode\rlap{[#1]}\kern2em}
\def\aut#1{{\rm#1}}                     \let\author        =\aut
\def\tit#1{{\sl#1}}                     \let\title          =\tit
\def\yea#1{{\rm(#1)}}   
\def\pub#1{{\rm#1}}                     
\def\jou#1{{\rm#1}}                     
\def\vol#1{{\bf#1}}                     
\def\pag#1{{\rm#1}}                     
\def\pla#1{{\rm#1}}                     

\def\nl{\hfill\break}                   
\let\lin\relax

\ref{1}
 \aut{Lodder J.J.},
 \lin\jou{Physica}
 \vol{116A},
 \yea{1982},
 \pag{45, 59, 380, 392},
 \jou{Physica}
 \vol{132A},
 \yea{1985},
 \pag{318}.

\ref{2}
 \aut{Lodder J.J.}, 
 \lin\jou{Physica}
 \vol{120A},
 \yea{1983}, \pag{1, 30, 566, 579}.

\ref{3}
 \aut{Lodder J.J.},
 \tit{Towards a Symmetrical Theory  of Generalised Functions},
 \jou{CWI tract }\vol{79},
 \yea{1991},
 \pub{CWI, Amsterdam}.

\ref{4}
\aut{Donoghue, J. F., Golowich E., Holstein B. R.},
\nl
\tit{Dynamics of the standard model},
\pub{Cambridge University Press},
\yea{1992}.

\ref5
 \aut{Lodder J. J.},
 hep-ph/9606306.

\ref6 
\aut{Muta, T.},
\title{Foundations of Quantum Chromodynamics},
\pub{World Scientific},
\pla{Singapore},
\yea{1987}.

\ref7 
\aut{Griffith, D.},
\title{Introduction to Elementary Particle Physics},
\pub{John Wiley \& Sons},
\pla{New York},
\yea{1987}.

\ref 8 
\aut{Gehrmann, T.},
hep-ph/9603380.

\ref 9 
\aut{Kundu, A.},
hep-ph/9504417.

\ref{10}  
 \aut{CDF collaboration},
 \jou{FERMILAB-Pub-96-020-E},
 \yea{1996}.

\endgroup
\bye